\documentclass[conference]{IEEEtran}
\IEEEoverridecommandlockouts
\usepackage{cite}
\usepackage{amsmath,amssymb,amsfonts}
\usepackage{algorithmic}
\usepackage{graphicx}
\usepackage{textcomp}
\usepackage[table]{xcolor}
\def\BibTeX{{\rm B\kern-.05em{\sc i\kern-.025em b}\kern-.08em
    T\kern-.1667em\lower.7ex\hbox{E}\kern-.125emX}}

\usepackage{enumitem}
\usepackage{booktabs}
\usepackage{subcaption}
\usepackage{multirow}
\graphicspath{{./images/}}

\definecolor{Gray}{gray}{0.95}
\newcolumntype{g}{>{\columncolor{Gray}}c}    
\begin{document}

\title{Time series signal recovery methods: comparative study
\thanks{© 20XX IEEE.  Personal use of this material is permitted.  Permission from IEEE must be obtained for all other uses, in any current or future media, including reprinting/republishing this material for advertising or promotional purposes, creating new collective works, for resale or redistribution to servers or lists, or reuse of any copyrighted component of this work in other works.}
}

\author{\IEEEauthorblockN{Firuz Kamalov}
\IEEEauthorblockA{\textit{Department of Electrical Engineering} \\
\textit{Canadian University Dubai}\\
Dubai, UAE\\
firuz@cud.ac.ae}
\and
\IEEEauthorblockN{Hana Sulieman}
\IEEEauthorblockA{\textit{Department of Mathematics and Statistics} \\
\textit{American University of Sharjah}\\
Sharjah, UAE \\
hsulieman@aus.edu}
}

\maketitle

\begin{abstract}
Signal data often contains missing values. Effective replacement (imputation) of the missing values can have significant positive effects on processing the signal. In this paper, we compare three commonly employed methods for estimating missing values in time series data: forward fill, backward fill, and mean fill. We carry out a large scale experimental analysis using 3,600 AR(1)-based simulated time series to determine the optimal method for estimating missing values. The results of the numerical experiments show that the forward and backward fill methods are better suited for times series with large positive correlations, while the mean fill method is better suited for times series with low or negative correlations. The extensive and exhaustive nature of the numerical experiments provides a definitive answer to the comparison of the three imputation methods.
\end{abstract}

\begin{IEEEkeywords}
time series, filling methods, imputation, AR, PACF, autoregression
\end{IEEEkeywords}

\section{Introduction}
One of the common issues that plagues real life data is missing values. The problem of missing data is present in many applications including signal processing, finance, meteorology, medicine, and others.  Errors in collection and recording can lead to incomplete data which can lead to false results. Analysis and conclusions derived based on incomplete data can be misleading. Therefore, the development of effective solutions for dealing with incomplete data is an important area of research. In this paper, we focus on handling incomplete data in the context of time series. The process of replacing the missing values is commonly referred to as imputation.  Dealing with time series data poses its unique challenges. Our goal is to compare various approaches to fill (impute) missing time series values and identify the most effective solution.

In certain cases simply ignoring the missing values may be acceptable. It is the easiest and most convenient approach that is frequently utilized by data scientist. However, a more nuanced approach is required when dealing with time series data. 
Temporal relationships are affected when data is dropped from a time series. For instance, seasonal patterns can be obfuscated when missing time series values are ignored. Stochastic models such as autoregressive and moving average assume a continuous sequence of ordered values and can produce delusive results in case of dropped values. As a result, missing time series data requires replacement. 

There exists a number of approaches to deal with incomplete data \cite{Pratama}. 
In the present study, we consider three of the most commonly employed methods for estimating missing values in time series: forward fill, backward fill, and mean fill.  In the \textit{forward} fill method, the missing values are replaced with the preceding observations, that is, if $x_t$ is missing then it is replaced with $x_{t-1}$.  In the \textit{backward} fill method, the missing values are replaced with the following observations, that is, if $x_t$ is missing then it is replaced with $x_{t+1}$. In the \textit{mean} fill method, the missing values are replaced with the average value of the sample series. The forward and backward fill methods have the advantage in scenarios where there is a strong positive correlation between the time series values whereas mean fill yields better performance in more volatile data. The three imputation methods are simple and easy to implement making them a very popular tool among the practitioners. Unlike the more exotic imputation techniques that require significant time and effort and do not necessarily produce optimal results, the classic filling methods considered in our study provide a fast and reliable avenue for replacing missing times series values. As a result, they remain popular and highly relevant. Our paper presents an exhaustive study of the methods' performance under a range of scenarios. 
We carry out a large scale numerical analysis based on 3,600 simulated time series to determine the most effective replacement technique.

In this paper, we focus on the autoregressive (AR) process of order 1. It is the most commonly used time series model. The results of the AR(1) process can be suitably extended to the general family of AR($p$) processes. Recall that the AR(1) process is given by the equation
\begin{equation}
\label{ar}
x_t = \phi x_{t-1} +w_t,
\end{equation}
where $x_t$ is the value of the time series at time $t$ and $w_t$ is white noise.
In our experiments, we simulate the AR(1) process and artificially remove a portion of the values. Then we reconstruct the missing values using the three filling techniques mentioned above. Our goal is to identify the filling method that produces sample partial autocorrelation function (PACF) values that are closest to the theoretical  PACF values. Our study is based on a total of 3,600 simulated time series using different mode coefficients. The results of the experiments show that mean fill outperforms the other methods.

Time series forecasting plays an important role in a number of applications including finance, medicine, engineering and others \cite{Hamilton, Kamalov1, Kamalov2}. The issue of missing times series values can lead to inaccurate forecasts. Therefore, a good understanding of imputation methods is crucial. We believe that our study would serve as a useful reference to interested parties both inside and outside of academia.
\section{Literature}
There exists a number of methods for dealing with incomplete time series data. An overview of the classic  imputation methods along with their advantages and disadvantages can be found in \cite{Pratama}. Imputation methods and algorithms are extensively implemented in statistical packages such as R \cite{Moritz}. A recent trend in imputation research involves the use of neural networks to estimate the missing time series values. For instance, the authors in \cite{Luo} employ generative adversarial networks whereas in \cite{Susanti} the authors use Bayesian networks to impute missing values in multivariate time series. In \cite{Fortuin}, the authors propose a deep sequential latent variable model to address the issue of interpretability. The authors employ non-linear dimensionality reduction using a VAE approach together with structured variational approximation. 

Domain-specific imputation methods have also been proposed. In \cite{Li}, the authors employ a multiview learning method to replace missing values in traffic-related time series data.  The model combines LSTM and SVR algorithms together with collaborative filtering techniques. The proposed method is able to account for the local and global variation in temporal and spatial views to capture more information from the existing data. The issue of missing data in transportation time series was also addressed in \cite{Sun}. The authors utilized an improved kNN-based imputation method to improve the accuracy by 34\%. The authors in \cite{Bokde} modify the pattern sequence algorithm to to simultaneously forecast and backcast missing values for imputation. Although test results were positive more extensive experiments are required to confirm the efficacy of the algorithm. Imputation methods have also been proposed in other domain such estimating forest biomass \cite{Nguyen} and water level forecasting \cite{Yang}.
A bagging algorithm to improve the existing imputation methods was proposed in \cite{Andiojaya}. The test results show that bagging has a positive effect on the performance of the imputation algorithms.

\section{Numerical experiments}
In this section, we carry out a range of numerical experiments to determine the best technique to estimate missing values in time series. Our focus is on the AR(1) based time series. We compare three approaches: forward fill, backward fill, and mean fill. The results of the experiments reveal that backward fill produces the most accurate results among all the tested methods.

\subsection{Methodology}
To analyze the performances of the filling models we measure their accuracy on AR(1) generated time series. To this end, we consider a range of AR(1) model coefficients $\phi$:  $0.1, 0.2, 0.3, 0.4$, $0.5, 0.6, 0.7, 0.8, 0.9$. 
For each value of $\phi$, we generate 100 different time series. Next, for each time series, we randomly drop a portion of the values. The resulting time series simulates a real life scenario of incomplete data. Then we estimate and replace the missing values using the above filling techniques. The sample PACF values are calculated for each restored time series. The filling methods are evaluated based on the difference between the theoretical PACF of the original process and the sample PACF based on the restored time series. For each value of $\phi$, we aggregate the results of the experiments on 100 simulated time series. We report the overall mean of the difference between theoretical and sample PACF values.
One of the key factors in the analysis of incomplete data is the quantity of missing values. In our experiments, we study the performance of filling algorithms under different drop rates: 5\%, 10\%, 20\%, and 30\% of the series values. As a result, we obtain a comprehensive perspective of the model performance.
A summary of the methodology is provided below.
\\
\\
\textbf{Algorithm}
\\
\line(1,0){150}
\\
For a fixed value of $\phi$, dropout rate $p$, and filling method $\mathcal{M}$.
\begin{enumerate}[label=\arabic*., itemsep=1ex]

\item Randomly generate 100 different sample time series based on the AR(1) model (Eq. \ref{ar}).
\item For each time series $x_t$, calculate the accuracy score using the following steps
\begin{enumerate}[label=\roman*., itemsep=1ex]
\item Drop a fraction $p$ of the time series values.
\item Replace the missing values using the method $\mathcal{M}$.
\item Calculate the sample PACF at lag $h=1$ based on the restored time series.
\item Calculate the accuracy score as the difference between the sample PACF and the theoretical PACF.
\end{enumerate}
\item Calculate the mean and standard deviation of accuracy scores over the 100 sample time series.
\end{enumerate}

The numerical experiments are implemented in Python using the \texttt{statsmodels} package \cite{Seabold}. We used Google Colab to carry out the experiments.

\subsection{Numerical experiments}

A range of numerical experiments with various parameter settings and dropout rates were performed to measure the performance of the filling methods. A total of 3,600 simulated sample time series were analyzed to obtain the final results.

\begin{figure}[!htb]
\begin{subfigure}{0.5\textwidth}
  \centering
 \includegraphics[clip, trim=5cm 0cm 3.5cm 0cm, width=1.0\textwidth]{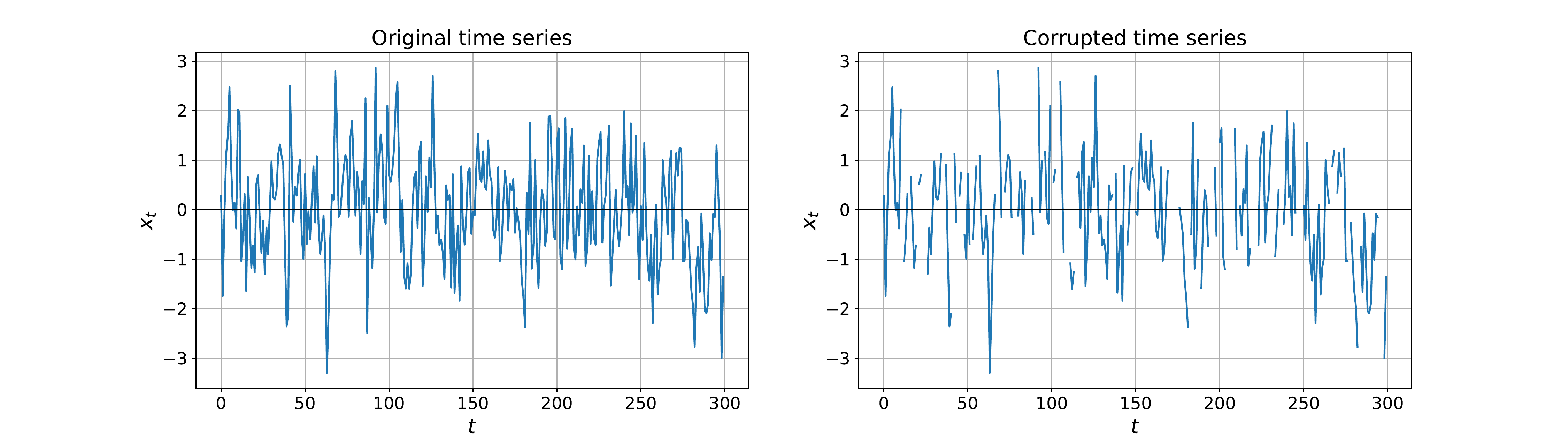}
  \caption{The original and corrupted times series. The corrupted series is missing 20\% of the original values. The original sample PACF at lag 1 is 0.329.}
  \label{fig:a}
\end{subfigure}
\newline
\begin{subfigure}{0.5\textwidth}
  \centering
  \includegraphics[clip, trim=5cm 0cm 3.5cm 0cm, width=\textwidth]{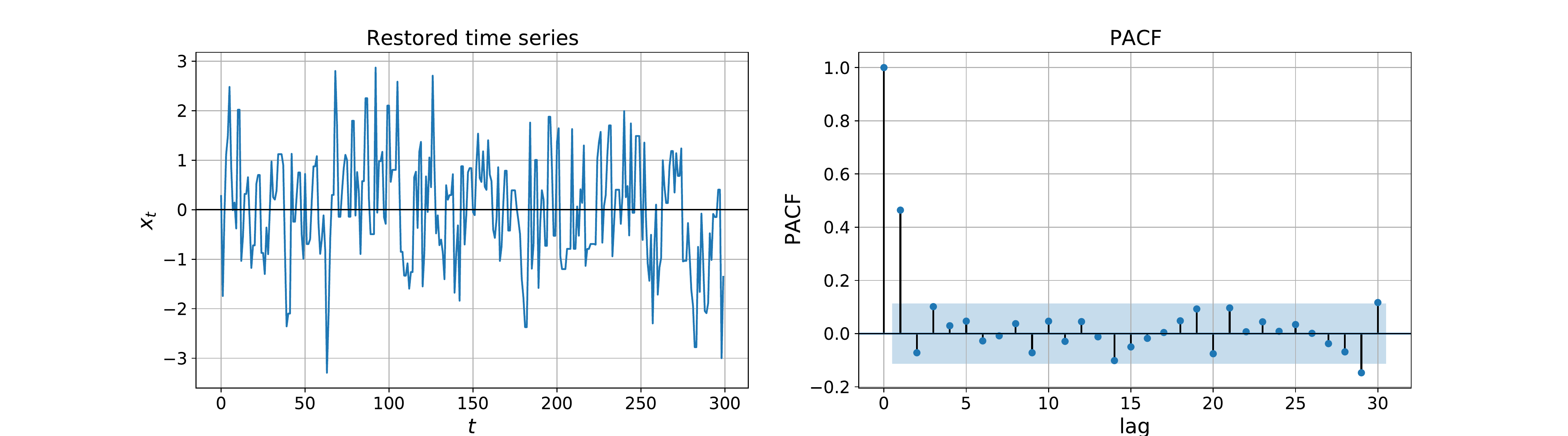}  
  \caption{The restored times series using forward fill method and the corresponding sample PACF  plot (0.464).}
  \label{forward}
\end{subfigure}
\newline
\begin{subfigure}{0.5\textwidth}
  \centering
  \includegraphics[clip, trim=5cm 0cm 3.5cm 0cm, width=1\textwidth]{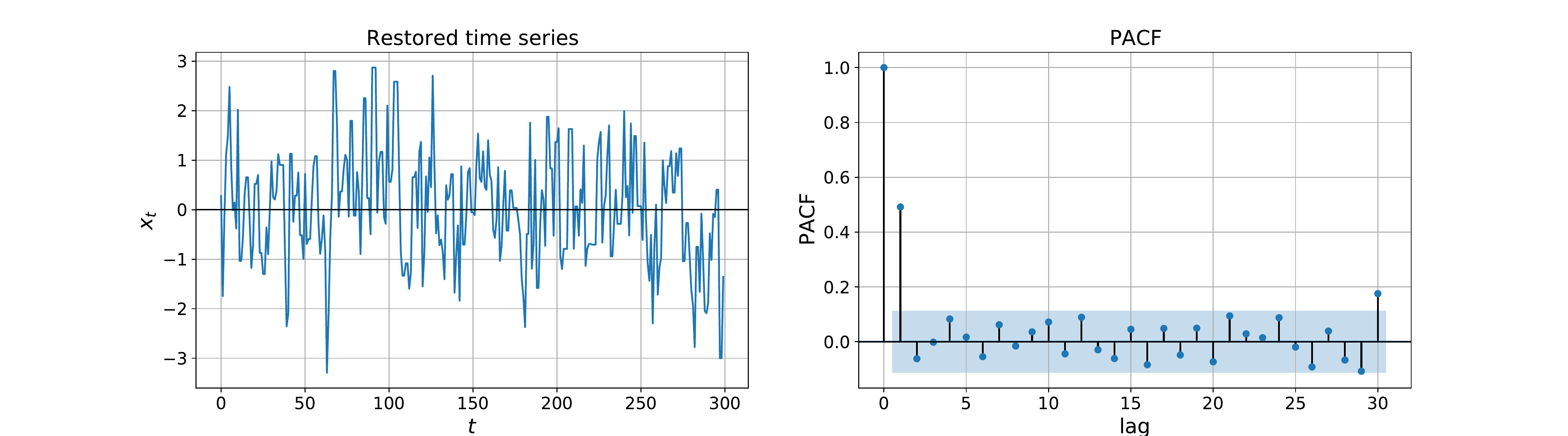}  
  \caption{The restored times series using backward fill method and the corresponding sample PACF plot (0.492) .}
  \label{backward}
\end{subfigure}
\newline
\begin{subfigure}{0.5\textwidth}
  \centering
  \includegraphics[clip, trim=5cm 0cm 3.5cm 0cm, width=1\textwidth]{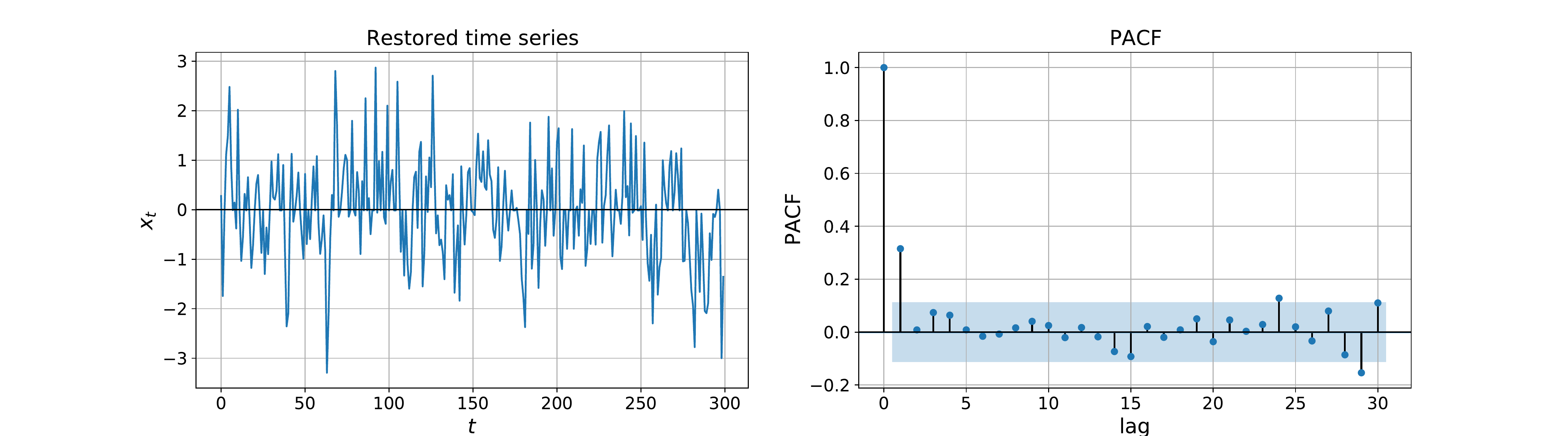}  
  \caption{The restored times series using mean fill method and the corresponding sample PACF plot (0.315).}
  \label{mean}
\end{subfigure}
\caption{The original sample time series is generated based on the AR(1) process with coefficient $\phi=0.4$. Then 20\% of the time series values were dropped to obtain an incomplete sample. The three filling methods are applied to restore the series.}
\label{example}
\end{figure}

To illustrate the experiments consider a single sample time series generated using the AR(1) process with parameter $\phi=0.4$ (Figure \ref{fig:a}). Note that in this case the theoretical PACF  is equal to 0.329. 
To mimic real-life incomplete data, we drop 20\% of the time series values as shown in Figure \ref{fig:a}. Then the filling methods are used to replace the missing values. As shown in Figures \ref{forward}-\ref{mean}, the restored series are close to the original sample time series. For each filling method, we compute the sample PACF (Figures \ref{forward}-\ref{mean}). To measure the performance of a filling method we take the absolute difference between the theoretical PACF (0.329) and the sample PACF obtained from the restored time series. Thus, the accuracy score of the forward fill method is $|0.329 - 0.464|=0.135$ (Figure \ref{forward}). The accuracy scores of the backward and mean fill methods are $0.163$ and $0.014$ respectively. This experiment is repeated 100 times for each combination of the model parameter $\phi$ and dropout rate values.

%

\begin{figure}[!htb]
  \centering
  \includegraphics[width=0.5\textwidth]{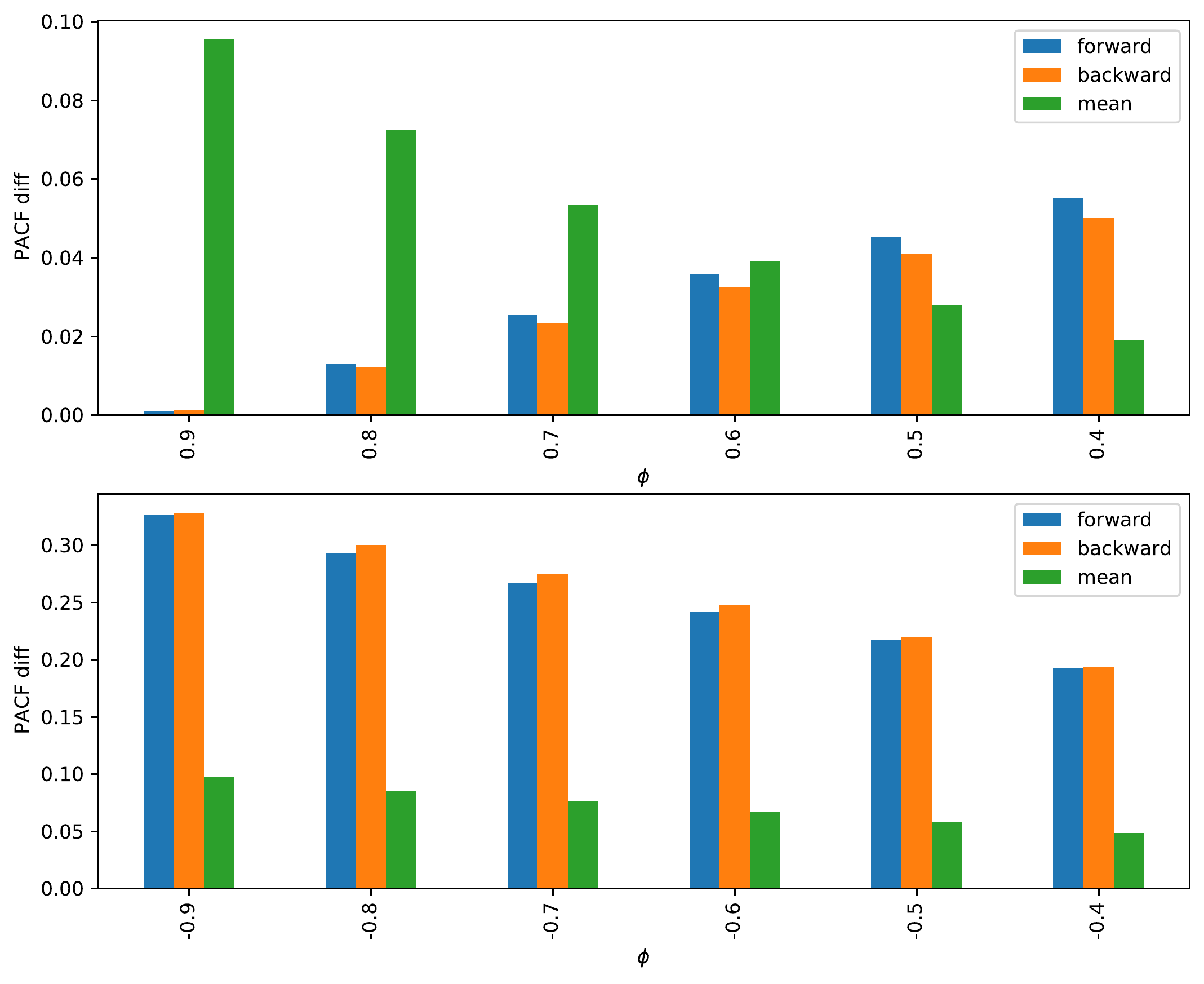}  
  \caption{The average true and estimated PACF difference for times series with 10\% missing values.}
  \label{drop10}
\end{figure}

\begin{figure}[!htb]
  \centering
  \includegraphics[width=0.5\textwidth]{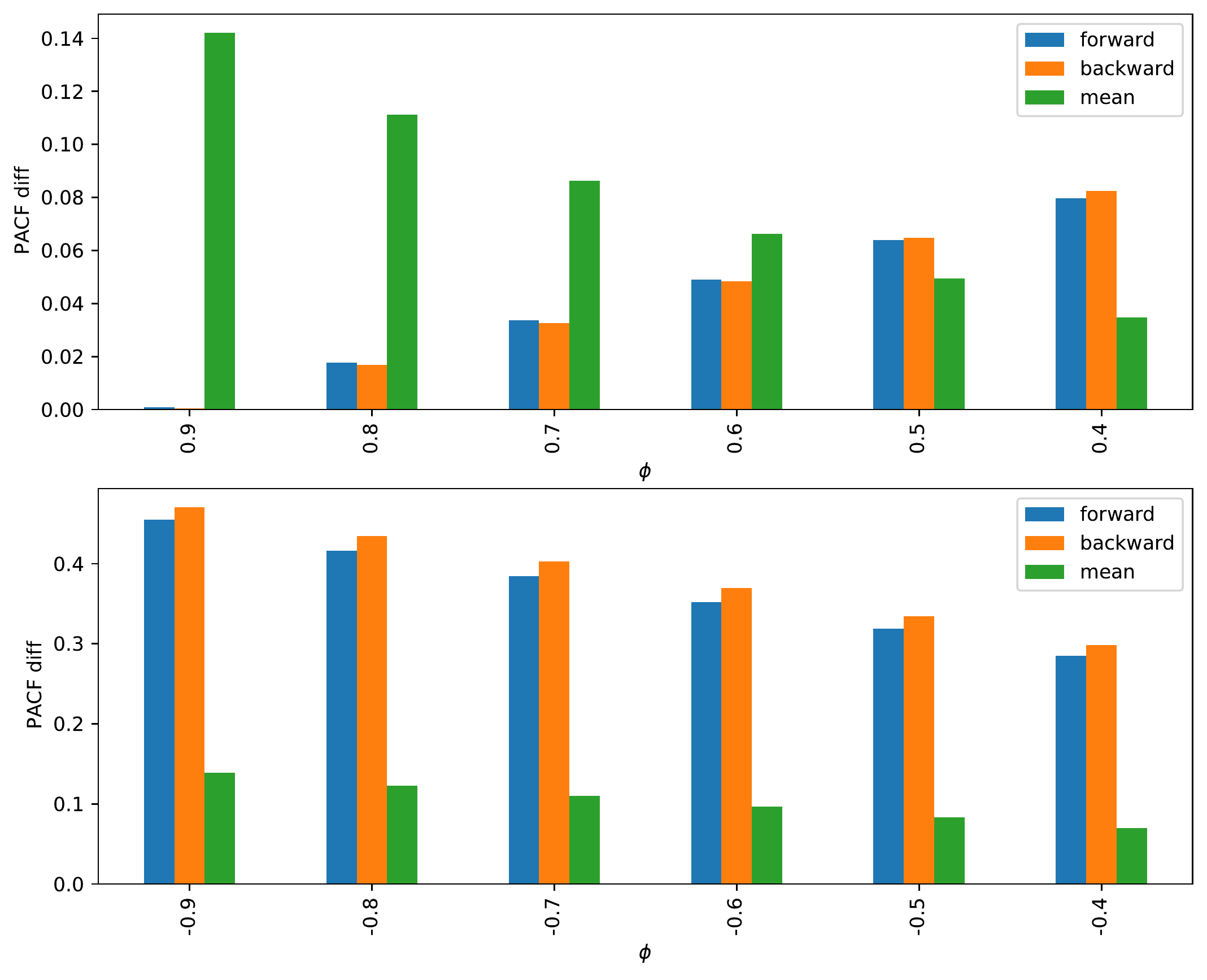}  
  \caption{The average true and estimated PACF difference for times series with 15\% missing values.}
  \label{drop15}
\end{figure}

\begin{figure}
  \centering
  \includegraphics[width=0.5\textwidth]{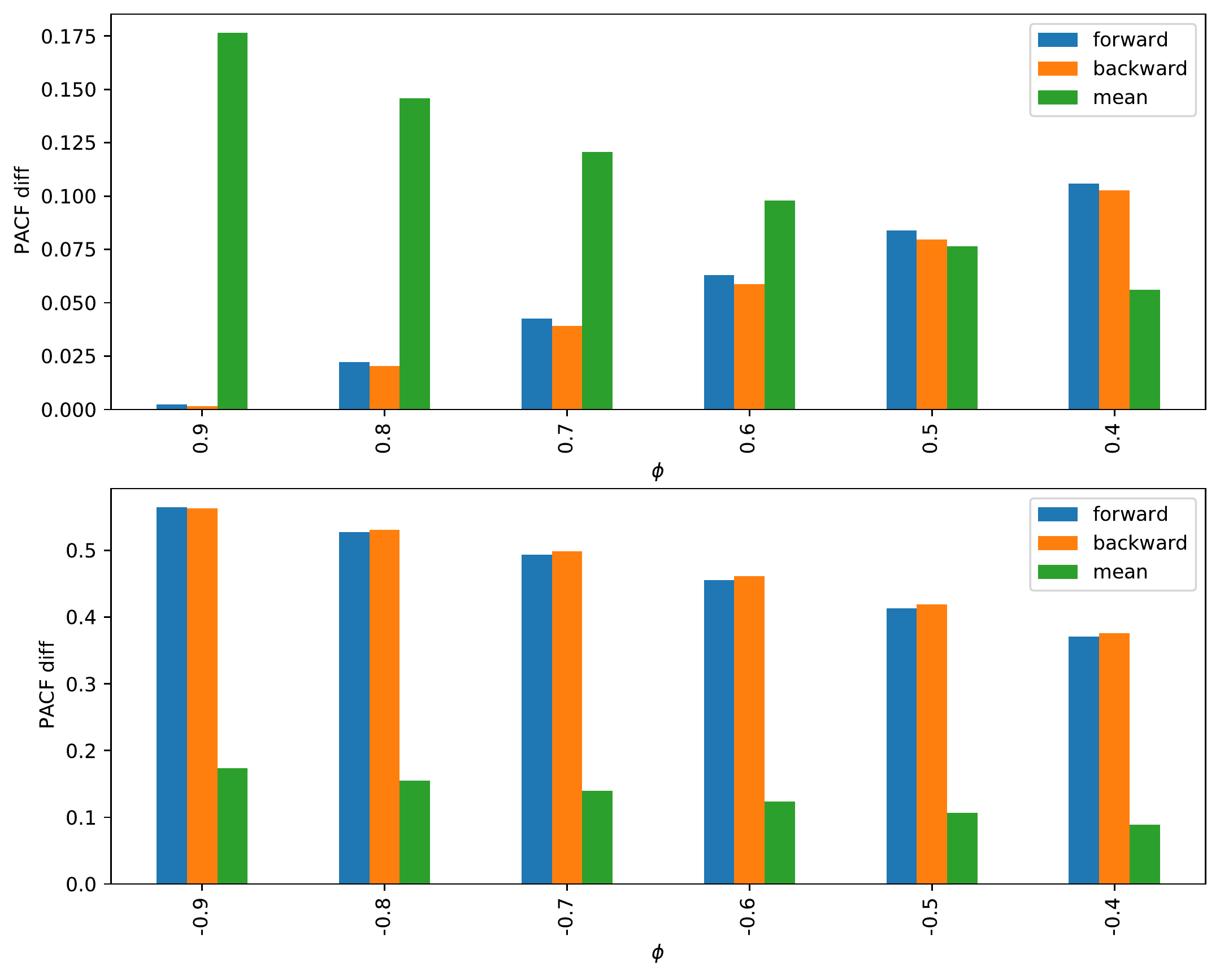}  
  \caption{The average true and estimated PACF difference for times series with 20\% missing values.}
  \label{drop20}
\end{figure}

\begin{figure}
  \centering
  \includegraphics[width=0.5\textwidth]{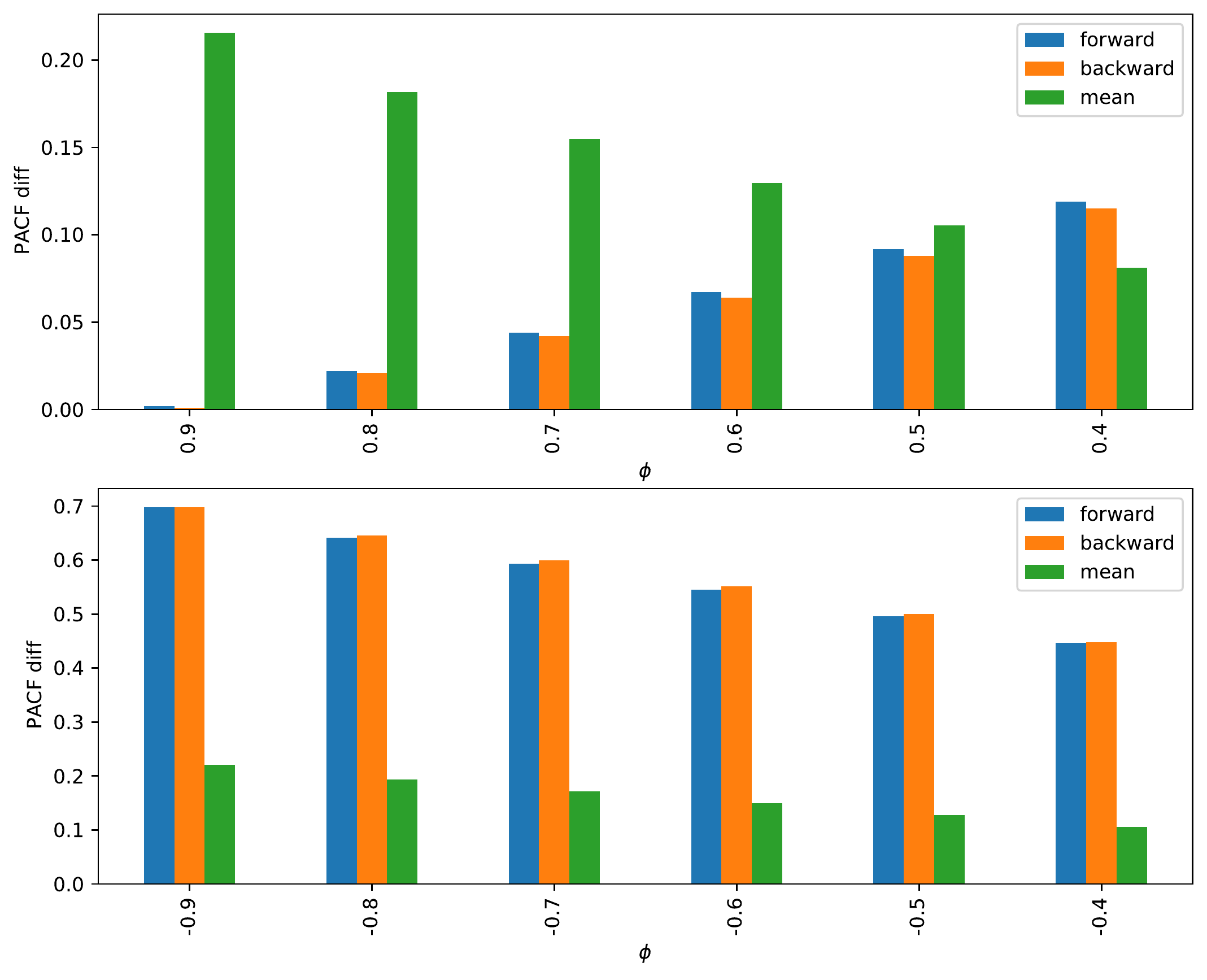}  
  \caption{The average true and estimated PACF difference for times series with 25\% missing values.}
  \label{drop25}
\end{figure}

The results of our numerical experiments are reported according dropout rates. Concretely, for each dropout rate, we repeat the experiment described in Figure \ref{example} 100 times for each value of $\phi$ and calculate the average difference between the theoretical and restored PACF. The results of the experiments are presented in Figures \ref{drop10}-\ref{drop25}. First, we consider the case when 10\% of the time series values are missing. As shown in Figure \ref{drop10}, the forward and backward fill methods produce small  errors (difference) at large positive values of $\phi$. On the other hand, mean fill produces relatively larger errors. Note that at large positive values of $\phi$ the time series has a strong trend. Therefore, the forward and backward methods that follow the the trend produce better results.
However, the performance of mean fill improves as the value of $\phi$ decreases. Eventually, beyond the value of $\phi=0.5$ , mean fill produces significantly better results than the forward and backward fill methods. At the lower values of $\phi$, the time series moves more sporadically with frequent changes in the direction of movement. As a result, the forward and backward fill methods overestimate the time series values. The mean fill method produces more conservative estimates that are less likely to miss the true time series value by a large margin.
When evaluating the performances of the filling methods across the range of values of $\phi$ mean fill yields undoubtedly better results. 

The performance of the filling methods on data with higher dropout rate is consistent with that of 10\% dropout rate. As shown in Figures \ref{drop15}-\ref{drop25}, the forward and backward fill methods perform well at large positive values of $\phi$. On the other hand, mean fill produces better results as the value of $\phi$ decreases. Concretely,  mean fill outperforms the other methods for all values $\phi \leq 0.5$. It is interesting to observe that while the forward and backward fill methods achieve a better performance with positive values of $\phi$, mean fill achieves better results with negative values of $\phi$. We also note that the performances of all three methods deteriorate as the dropout rate increases. For instance, the PACF error of mean fill increases from under 0.10 to over 0.20 at $\phi=0.9$ as the dropout rate increase from 10\% to 25\%. Similarly, the PACF error of forward fill increases from under 0.35 to almost 0.70 at $\phi=-0.9$ as the dropout rate increase from 10\% to 25\% (Figures \ref{drop10}-\ref{drop25}).

\section{Conclusion}
In this paper, we analyzed the performance of three methods to fill the missing values in a time series data. Concretely, we studied the performance of the forward, backward, and mean fill methods in restoring the missing values from AR(1) generated time series sample. We carried out a total of 3,600 simulations with a range of dropout rates and model parameter values. The performance of the filling methods was measured based on the difference between the true and estimated PACF values. The results of the numerical experiments show that mean fill significantly outperforms the other methods at values $\phi\leq 0.5$. The results are consistent across different dropout rates. Forward and backward fill  achieve better results at large positive values of $\phi$. The performance of forward fill is due to the positive trend in the time series for positive values of $\phi$. As a result, forward fill achieves accurate results by following the trend estimates. Conversely, mean fill achieves better results for value $\phi \leq 0.5$ which is explained by higher time series stochasticity. The mean fill method produces more conservative estimates that are better suited for frequently alternating series.

We conclude that the forward and backward fill methods are better suited for time series with strong positive correlations between its values. Mean fill is better suited for time series with low positive correlation or negative correlations between its values. The results hold across different dropout rates. Since the AR(1) process is one of the commonly encountered models, the outcomes of this study can be reasonably extended to other stochastic processes. Given the importance of time series forecasting \cite{Kamalov3, Kamalov4}, our study would be a valuable reference to both academics and industry practitioners.

\end{document}